%% file: main.tex
\documentclass{vgtc}                          % final (conference style)
\graphicspath{{figures/}{pictures/}{images/}{./}} % where to search for the images

\usepackage[dvipsnames,svgnames,x11names]{xcolor}

\usepackage{amsthm}
\usepackage{times}                     % we use Times as the main font
\usepackage{tabu}                      % only used for the table example
\usepackage{booktabs}                  % only used for the table example
\usepackage{lipsum}                    % used to generate placeholder text
\usepackage{mwe}                       % used to generate placeholder figures
\usepackage{enumitem}
\usepackage{amsmath}
\usepackage{amssymb}

\usepackage{colortbl}

\usepackage{amsthm}
\usepackage{changepage}

\definecolor{appleredlight}{RGB}{255, 105, 97}
\definecolor{appleorangelight}{RGB}{255, 179, 64}
\definecolor{appleyellowlight}{RGB}{255, 212, 38}
\definecolor{applegreenlight}{RGB}{48, 219, 91}
\definecolor{applemintlight}{RGB}{102, 212, 207}
\definecolor{appleteallight}{RGB}{93, 230, 255}
\definecolor{applecyanlight}{RGB}{112, 215, 255}
\definecolor{applebluelight}{RGB}{64, 156, 255}
\definecolor{appleindigolight}{RGB}{125, 122, 255}
\definecolor{applepurplelight}{RGB}{218, 143, 255}
\definecolor{applepinklight}{RGB}{255, 100, 130}
\definecolor{applebrownlight}{RGB}{181, 148, 105}

\definecolor{applerednormal}{RGB}{255, 69, 58}
\definecolor{appleorangenormal}{RGB}{255, 159, 10}
\definecolor{appleyellownormal}{RGB}{255, 214, 10}
\definecolor{applegreennormal}{RGB}{48, 209, 88}
\definecolor{applemintnormal}{RGB}{99, 230, 226}
\definecolor{appletealnormal}{RGB}{64, 200, 224}
\definecolor{applecyannormal}{RGB}{100, 210, 255}
\definecolor{applebluenormal}{RGB}{10, 132, 255}
\definecolor{appleindigonormal}{RGB}{94, 92, 230}
\definecolor{applepurplenormal}{RGB}{191, 90, 242}
\definecolor{applepinknormal}{RGB}{255, 55, 95}
\definecolor{applebrownnormal}{RGB}{172, 142, 104}

\definecolor{applereddark}{RGB}{233, 21, 45}
\definecolor{appleorangedark}{RGB}{197, 83, 0}
\definecolor{appleyellowdark}{RGB}{161, 106, 0}
\definecolor{applegreendark}{RGB}{0, 137, 50}
\definecolor{applemintdark}{RGB}{0, 133, 117}
\definecolor{appletealdark}{RGB}{0, 129, 152}
\definecolor{applecyandark}{RGB}{0, 126, 174}
\definecolor{applebluedark}{RGB}{30, 110, 244}
\definecolor{appleindigodark}{RGB}{86, 74, 222}
\definecolor{applepurpledark}{RGB}{176, 47, 194}
\definecolor{applepinkdark}{RGB}{231, 18, 77}
\definecolor{applebrowndark}{RGB}{149, 109, 81}

\usepackage{longfbox} % for drawing frames on stakeholder aliases

\makeatletter
\newdimen\@tempdimd
\makeatother

\newfboxstyle{patternparamrad}{padding-top=1pt, padding-bottom=1pt, padding-left=3pt, padding-right=3pt, border-style=solid, height=6.5pt, border-radius=4pt}

\newfboxstyle{patternparam}{padding-top=1pt, padding-bottom=1pt, padding-left=3pt, padding-right=3pt, border-style=none, height=4pt}

\usepackage{kotex}

\usepackage{mathptmx}                  % use matching math font

\onlineid{0}

\vgtccategory{Research}

\vgtcinsertpkg

\title{Understanding Bias in Perceiving Dimensionality Reduction Projections}

\author{Seoyoung Doh$^1$\thanks{e-mail: 0303dsy@snu.ac.kr} $\enspace$ Hyeon Jeon$^1$\thanks{e-mail: hj@hcil.snu.ac.kr}  $\enspace$ Sungbok Shin$^2$\thanks{e-mail: sungbok.shin@inria.fr} $\enspace$ Ghulam Jilani Quadri$^3$\thanks{e-mail: quadri@ou.edu} $\enspace$ Nam Wook Kim$^4$\thanks{e-mail: nam.wook.kim@bc.edu} $\enspace$ Jinwook Seo$^1$\thanks{e-mail: jseo@snu.ac.kr, corresponding author}\\ %
        \scriptsize $^1$Seoul National University $\quad$
     \scriptsize $^2$Inria, Université Paris-Saclay $\quad$
     \scriptsize $^3$University of Oklahoma $\quad$ 
     \scriptsize $^4$Boston College}

\teaser{
  \centering
  \includegraphics[width=\linewidth]{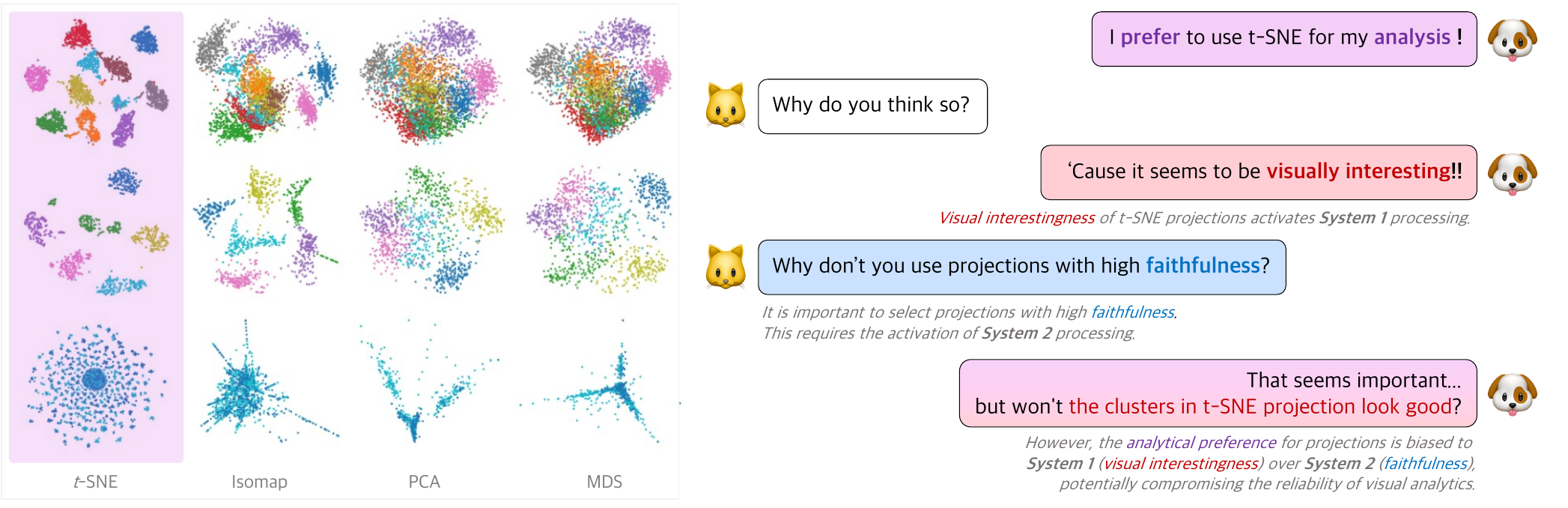}
  \vspace{-7mm}
  \caption{The illustration of our motivating scenario and its alignment with dual-system theory (\autoref{sec:dualsystem}).
  We assume a scenario where a practitioner wants to select DR techniques that produce projections suitable for analyzing their dataset. 
  It is crucial to select projections with high faithfulness for reliable analysis, but practitioners tend to be more biased towards the visual interestingness. We verify the existence of such bias and investigate its underlying causes, providing a grounded basis for mitigating the biases.}
  \label{fig:teaser}
}

\abstract{
    \input{sections/00_abstract}
} % end of abstract


\begin{document}


\firstsection{Introduction}

\maketitle

\input{sections/01_introduction}
\input{sections/03_background}

\input{sections/04_user_study}

\input{sections/05_results}
\input{sections/06_discussions}
\input{sections/07_conclusion}
\bibliographystyle{abbrv-doi-narrow}

\bibliography{ref}
\end{document}

%% file: sections/00_abstract.tex
% 중학교 2학년이 봐도 이해할 수 있을 정도로,,
% 중요한 내용을 제대로 담는 것!이 무엇보다 중요,, "contents-oriented" %
Selecting the dimensionality reduction technique that faithfully represents the structure is essential for reliable visual communication and analytics. 
In reality, however, practitioners favor projections for other attractions, such as aesthetics and visual saliency, over the projection's structural faithfulness, a bias we define as \textit{visual interestingness}. 
In this research, we conduct a user study that (1) verifies the existence of such bias and (2) explains why the bias exists. 
Our study suggests that visual interestingness biases practitioners' preferences when selecting projections for analysis, and this bias intensifies with color-encoded labels and shorter exposure time.
% immediate impressions of visual interestingness bias practitioners' preference in selecting projections to analyze. 
Based on our findings, we discuss strategies to mitigate bias in perceiving and interpreting DR projections. 

%% file: sections/01_introduction.tex
Dimensionality reduction (DR) is a popular technique for visually interpreting and analyzing high-dimensional data.
To visualize data with DR, practitioners should first select DR techniques that align with their analytical tasks. 
Here, \textit{faithfulness} \cite{nonato19tvcg, nguyen13pvis, hyeon25stopmisusing}, the degree to which the structural characteristics of the original data are preserved without distortions, should be prioritized to ensure that the projection reliably supports the tasks. 
However, analysts may prioritize other factors, such as aesthetics or visual saliency, preferring projections with cleanly separated clusters and visually distinct boundaries (\autoref{fig:teaser}).
% They can also choose projections having structural characteristics that align with their expectations \cite{hyeon25stopmisusing}. %visual interestingness 
%over faithfulness. 

This bias towards \textit{visual interestingness} \cite{seo05infovis, yunhai18tvcg, friedman87}, the degree to which projections exhibit perceptually appealing patterns,  may prompt analysts to select unsuitable techniques. 
This issue is particularly concerning for researchers and data analysts who regularly perform visual analytics with high-dimensional datasets but have limited DR literacy, as their vulnerability to the bias threatens the reliability of their scientific discoveries \cite{cashman25arxiv}.
% For example, analysts can inadvertently adopt t‑SNE or UMAP for distance projection tasks, even though these methods exaggerate cluster separation at the cost of distorting distance relationships~\cite{hyeon25stopmisusing}.
%potentially degrading analytical robustness and producing misleading interpretations.

% """In reality, however, practitioners favor projections for other attractions, such as aesthetics and visual saliency, over the projection’s structural faithfulness, a bias we define as visual interestingness."""
% 하지만 현실에서, practitioner들은 faithfulness보다 visual attraction을 더 중요시 하는데, 우리는 이를 ‘visual interestingness’라는 편향으로 정의한다.
%However, in real-world settings, 
Our work aims to empirically investigate the impact of visual interestingness on analytical preference of DR projections in practice. 
For this, we address the following research questions:
\begin{itemize}[itemsep=0.01cm, topsep=0.05cm, leftmargin=15pt]
    \item (\textbf{RQ1}) Do practitioners tend to make biased selections by favoring visual interestingness of projections over faithfulness?
    \item (\textbf{RQ2}) If such bias is observed, why does such bias occur?
\end{itemize}
We conduct a two-phase user study for the purpose.
In the first phase, we obtain the ranking of DR projections based on visual interestingness by asking the participants. 
In the second phase, we present participants with the same projections alongside artificially created faithfulness scores that contradict the visual interestingness rankings from phase 1. 
By doing so, we examine which factor participants prioritize when selecting projections for analysis.
% in situations where visual appeal and faithfulness metrics conflict. 
% We end our study by conducting post-hoc interview, to analyze qualitative insights of participants' including whether they are aware of biased selection.
% We also conduct post-hoc interviews to gather qualitative insights into participants' decision-making processes, including their awareness towards bias or the visual patterns that incur such bias.

% """Our study suggests that immediate impressions of visual interestingness bias practitioners’ preference in selecting projections to analyze. """
% 우리의 스터디에서 밝히기를, visual interestingness의 즉각적인 인상이 분석을 위한 Projection을 고르는 데 있어 practitioner들의 preference를 bias한다.
% 그리고 우리는 유저스터디에서, practitioner들이 실제로 biased selection을 하고 있음을 관찰할 수 있었다. 사람들은 analytical preference에 부합하는 projection을 선택할 시 visual interestingness를 기준으로 선택하는 경향성을 매우 강하게 보였다. 이러한 visual 기반의 즉각적 선택은 monochrome projection일 경우 cluster separation이나 structure가 강할 때, polychrome projection일 경우 class separation이 강할 때 더욱 두드러진다. 또한 시간이 짧아질수록 전반적으로 이러한 visual cue 기반 선택의 경향이 더 강화되는 경향성도 보였다. 특히 이들은 faithfulness를 보여주는 metric을 거의 고려하지 않았는데, 이는 DR의 literacy와는 무관하게 나타난 양상이었다. 다시 말해 이들은 고차원 데이터의 구조 보존 및 왜곡 여부와 무관하게, 당장에 시각적으로 attractive한 projection을 선택하였다.
Our results verify the bias towards visual interestingness over faithfulness in determining analytical preference of DR projections, i.e., the degree to which analysts prefer to use projections for their analysis. 
% selecting DR projection for analysis. 
% that practitioners make biased selections. 
When selecting projections, participants exhibit a strong tendency to rely on visual interestingness, which is particularly pronounced with color-encoded class labels.
We also find that the bias intensifies with clear class or cluster boundaries. 
These findings prompt discussions on design strategies that can help mitigate bias in perceiving DR projections, enabling more reliable use of DR in visual analytics and communications. 

% when the projection displays strong cluster separation or structure in the monochrome condition, strong class separation in the polychrome condition, and also under shorter time constraints.
%This visually driven, intuitive decision-making is particularly prominent when the projection displays strong cluster separation or structure in the monochrome condition, and strong class separation in the polychrome condition. 
% Moreover, this tendency becomes even more pronounced under shorter time constraints.%regardless of their level of DR literacy. 
%Notably, participants rarely consider the faithfulness metrics provided, regardless of their level of DR literacy. 
%In other words, they consistently select visually attractive projections, even when such projections do not accurately preserve the structure of the original high-dimensional data.
% For RQ2, practitioners are largely unaware that they are making biased selections, which are, here too, agnostic to practitioners’ DR literacy levels.
%In fact, some participants do not even agree that such selections should be considered a form of `bias'.
%
%

% % In summary, …
% % % % % Takeaway
In summary, our contributions are as follows:
\begin{itemize}[itemsep=0.01cm, topsep=0.05cm, leftmargin=15pt]
    \item We identify \textbf{perceptual bias} in selecting DR projections to analyze towards visual interestingness over faithfulness.
    \item We empirically verify the existence of such bias and investigate why such bias occurs through a \textbf{user study} with 32 participants. 
    \item We recommend \textbf{strategies} to \textbf{mitigate} bias in perceiving DR projections. 
    % analyAnalysis of the bias using cognitive psychology-based principles (e.g., Dual-System Theory)%grounded in Dual-System Theory.
    % \item Establishment of a foundation for developing mitigation strategies through identification of features that influence visual interestingness.
    % visual interestingness에 영향을 주는 feature들을 알아봄으로써, 이 문제를 해결하기 위한 mitigation strategy의 초석을 다짐
\end{itemize}

%% file: sections/03_background.tex
\section{Preliminaries}
% % 우리는 우선 DR과, 우리의 웤에서 주요하게 다뤄지는 두 가지 요소; (1) faithfulness와, (2) visual interestingness 를 설명한다. 그러고는 이 요소들과 align?되는 인지심리학적 이론, Dual System에 대해 설명한다.

% Our work is relevant to two main areas of literature: (1) DR , and (2) the dual system theory. 
Our work draws upon two main areas of literature: (1) DR and (2) dual-system theory from cognitive psychology.

% 3.1. Dimensionality Reduction
% (1) Faithfulness, (2) Visual Interestingness
% 3.2. Dual-System Theory
% DR보다 더 많은 공간과 내용을 할애할 것..

\subsection{Dimensionality Reduction}
% \subsection{Faithfulness of Dimensionality Reduction}
% We first review DR, then discuss the importance of assessing the faithfulness, and visual interestingness of DR projections.
We first review DR, then discuss the importance of assessing both the faithfulness and visual interestingness of DR projections.

% \noindent\textbf{Dimensionality reduction.}
% DR techniques receive high-dimensional data as input and produce a low-dimensional representation that preserves the original structural characteristics \cite{nonato19tvcg, cashman25arxiv, carreira1997review}, where the output dimension is typically two or three. By doing so, DR provides a visual summarization of high-dimensional data, depicting characteristics such as class relationships or cluster structures \cite{kwon18tvcg, jeon24tvcg}.

% However, even with the same dataset, DR projections can show significantly different visual patterns depending on the choice of technique and hyperparameters \cite{nonato19tvcg, wattenberg2016how, appleby22cgf}. 
% Thus, when selecting a DR projection, analysts must choose both the appropriate technique and configure its hyperparameters in accordance with their intended analytical task \cite{wattenberg2016how, hyeon25stopmisusing}. 

\vspace{2.5pt}
\noindent\textbf{Dimensionality reduction.}
DR techniques receive high-dimensional data as input and transform it into low-dimensional (typically two or three) representations that preserve structural characteristics \cite{nonato19tvcg, cashman25arxiv}. 
By doing so, analysts can visually investigate data characteristics such as relationships between different classes \cite{kwon18tvcg, jeon24tvcg}.
However, even with the same dataset, DR projections can show different visual patterns depending on the chosen technique and hyperparameters \cite{nonato19tvcg, wattenberg2016how}. Therefore, analysts must carefully select both techniques and hyperparameters to align with their analytical objectives \cite{wattenberg2016how, hyeon25stopmisusing}.

\vspace{2.5pt}
\noindent\textbf{Faithfulness of DR projections.}
When selecting a DR technique, evaluating \textit{faithfulness}, the degree to which the original high-dimensional structure is preserved in the low-dimensional projection, is crucial for achieving reliable visual analytics \cite{hyeon25stopmisusing, nonato19tvcg}. 
Various metrics have thus been proposed to assess DR faithfulness \cite{hyeon23vis}. 
For example, \textit{trustworthiness and continuity} \cite{venna06nn} metrics evaluate how well neighborhood relationships are maintained in projections.

\vspace{2.5pt}
\noindent\textbf{Visual interestingness of DR projections.} 
Building on previous literature that defines interesting projections \cite{yunhai18tvcg, seo05infovis, friedman87}, we define visual interestingness as follows:
\vspace{4pt}
\begin{adjustwidth}{0.4cm}{0.4cm}
\noindent
\textbf{Definition.} \textit{Visual interestingness} of a projection denotes the degree to which the projection exhibits visually salient, distinctive, or appealing patterns that elicit perceptual attention.
\end{adjustwidth}
% captures and draws viewers' visual attention through factors other than the faithful representation of the data's structure, such as visual saliency, %\cite{shin23tvcg}
% distinctive patterns, %\cite{healey12tvcg}
% and aesthetic appeal. %\cite{he23tvcg}
\vspace{4pt}

\noindent
It is worth noting that our scope of visual interestingness does not cover cognitive interpretations of DR projections, such as engagement or alignment of patterns with analysts' expectations.

\vspace{2.5pt}
\noindent\textit{Our Contribution. }
We empirically verify that practitioners tend to prioritize visual interestingness over faithfulness when selecting DR projections for their analysis. Based on our findings, we discuss strategies to mitigate such bias.

\subsection{Dual-System Theory: System 1 and System 2}

\label{sec:dualsystem}

% 우리는 인간의 두 사고 과정을 인지심리학적으로 설명한 'dual system theory'를 소개하고, 이 이론이 DR projections selection의 상황, 그리고 우리의 실험 상황과 어떻게 align되는지 설명한다. 이를 통해 우리가 밝히고자 하는 bias in projection selection의 원인이 이 이론으로 설명될 수 있음을 밝힌다.
% We introduce a cognitive theory called `dual-system theory' and demonstrate how this theory aligns with scenarios of DR projection selection and our experimental design. Through this, we explain how bias in projection selection can be understood within this cognitive framework.

Our hypothesis and study design is grounded on \textit{dual-system theory}, a cognitive psychological framework that explains human thinking processes.
We first introduce the theory, then discuss how this theory aligns with our problem statement and the bias in perceiving DR projections. 

\vspace{2.5pt}
\noindent
\textbf{Dual-system theory.} 
% 
% 인간의 visualization perception을 분석함에 있어, 인지심리학 이론인 dual system이 자주 사용된다.
% Our study is grounded on a cognitive theoretical framework called .
% Dual System은 인간의 사고를 크게 두 가지의 시스템으로 구분한다; 직관적이지만 빠른 thinking을 담당하는 System 1과, 논리적이지만 느린 thinking을 담당하는 System 2가 있다. 이 둘은 서로 상호보완적인 관계이다; 가령 빠르고 직관적이지만 섣부른 판단을 내릴 우려가 있는 System 1을 느리지만 이성적인 System 2이 보완해준다. 또한 처해진 태스크에 따라 System 2의 게으른 thinking을 System 1이 효율화해주기도 한다.
\textit{Dual-system theory} \cite{kahneman11thinking, tversky74science} posits that human thinking operates through two processing systems: System 1, which is fast and intuitive, and System 2, which is slower but more logical and analytical. 
These two systems work together in a complementary manner. 
For instance, System 1 enables quick judgments but may lead to premature or biased decisions, which can be corrected by more deliberate reasoning of System 2. 
Conversely, in time-limited or routine tasks, System 1 can help optimize cognitive processing by compensating for System 2’s slower and more effortful thinking.

\vspace{2.5pt}
\noindent
\textbf{Alignment of the dual-system theory with our study.}
% 우리는 dual-system theory를 우리의 연구와 align시키는데, 우리는 system 1에 해당하는 요소로서 visual interestingness를, system 2에 해당하는 요소로서 faithfulness를 상정(정의)하였다. 그 이유는 system 1이 visual에 기반하여 직관적이고 빠른 processing을 한다고 알려져 있고 \cite{}, 이는 우리 연구에서 즉각적인 시각적 perception을 일으키는 visual interestingness와 부합한다고 할 수 있다. 
% 또한 system 2의 경우 객관적인 수치 데이터에 기반하여 느리지만 논리적인 processing을 한다고 알려져 있는데 \cite{}, 이는 우리 연구에서 reliability of DR technique depending on their task를 판단하는 numerical component인 faithfulness와 부합한다고 할 수 있다.
We align our study with dual-system theory by conceptualizing visual interestingness as a System 1 process and faithfulness as a System 2 process. 
% visual interestingness는 시각적 요소들을 통해 practitioner들의 인지를 순식간에 사로잡으므로, system 1의 fast and intuitive processing을 activate한다고 할 수 있다.
We map visual interestingness to System 1 as it captures practitioners' attention instantaneously through visual elements \cite{healey12tvcg}, thereby activating fast and intuitive processing. 
% The immediate visual appeal of well-separated clusters and clear color distinctions triggers automatic cognitive responses that bypass deliberate analysis.
% System 1 is characterized by fast, automatic, and intuitive processing based on visual perception \cite{}, which corresponds to the immediate visual appeal and clustering clarity that drives visual interestingness in our study. 
% 또한 faithfulness는 객관적 수치 요소들을 통해 practitioner들의 이성적 사고를 요구하므로, system 2의 slow but deliberate reasoning을 activate한다고 할 수 있다.
Faithfulness requires practitioners to engage in rational evaluation of objective numerical scores \cite{green08vast}; we thus align the faithfulness with System 2's deliberate reasoning that requires a longer time.
%% 카이페이퍼에 표를 넣읍습니다 표표표
% Faithfulness metrics demand systematic comparison of quantitative indicators to assess DR technique reliability, necessitating effortful cognitive processing. 
% Conversely, System 2 involves slower, deliberate, and analytical processing that relies on numerical evidence \cite{}, aligning with faithfulness metrics that quantitatively assess the reliability of DR techniques for specific analytical tasks. 
% This theoretical framework provides a cognitive foundation for understanding why practitioners may default to visual characteristics despite the availability of objective quality measures.

\vspace{2.5pt}
\noindent \textit{Our Contribution. }
% Building on this theoretical foundation, our experimental design examines how practitioners balance visual interestingness (System 1 engagement) and faithfulness considerations (System 2 engagement) when selecting DR projections. 
% 우리는 또한 시간이 짧을수록, 또 visual saliency가 클수록 System 1 작용이 강해진다는 기존 파인딩 \cite{} 과 얼라인해 exposure time과 color encoding of class label이 such bias에 어떤 영향을 미치는지 탐구했다. 
Building on this theoretical foundation, our experimental design examines System 1 and System 2 impacts the analytical preference of DR projections. We further align with established findings that shorter exposure times and higher visual saliency amplify System 1 processing \cite{healey12tvcg, morariu23tvcg, bibal16} by investigating how exposure duration and color encoding of class labels influence such bias. By controlling these factors, we identify empirical evidence of the cognitive mechanisms of perceiving and interpreting DR projections.

%% file: sections/04_user_study.tex
\begin{figure*}
    \centering
    \includegraphics[width=1\textwidth]{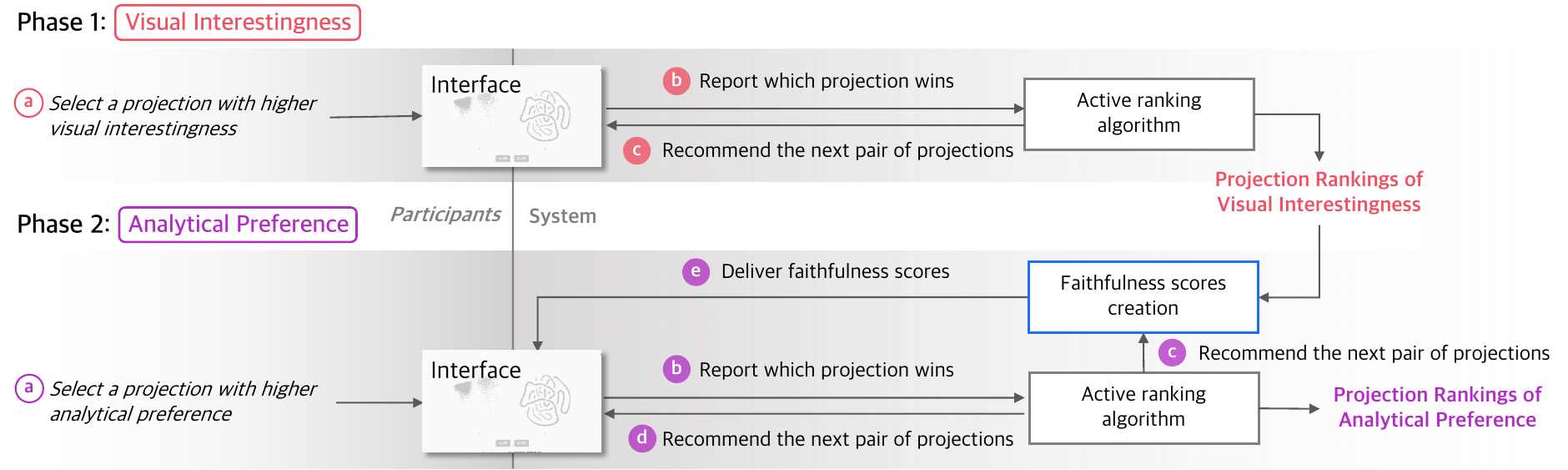}
    \vspace{-5mm}
    \caption{The illustration of our study design (\autoref{sec:userstudy}). In Phase 1, participants view pairs of projections and select the one with higher visual interestingness. Phase 2 repeats this process for analytical preference, but with artificially generated faithfulness metrics that favor less visually interesting projections. This design enables us to quantify participants' bias toward visual interestingness when it conflicts with faithfulness. Participant actions are denoted in italics, while system actions appear in regular text.}
    \label{fig:userstudy}
\end{figure*}

\section{User Study}

\label{sec:userstudy}

We detail our study, which reveals the existence of bias towards System 1 in perceiving DR projections (\autoref{fig:userstudy}).

\subsection{Objectives}

We aim to understand the interplay between three key components: (1) \textcolor{applereddark}{Visual interestingness} \cite{seo05infovis, yunhai18tvcg, friedman87}, which stands for the degree to which a projection captures an analyst’s visual attention, e.g., based on its visual salience, patterns, or aesthetic qualities, (2) \textcolor{applebluedark}{Faithfulness} \cite{nonato19tvcg, nguyen13pvis, hyeon25stopmisusing}, i.e., the degree to which projections accurately represent the original high-dimensional data without distortions, and (3) \textcolor{applepurpledark}{Analytical preference} \cite{xia22tvcg}, which is defined as the degree to which analysts want to use projections for their analysis.
To do so, we verify the following hypotheses:
\begin{itemize}[leftmargin=15pt, itemsep=0pt, topsep=0.05cm]
    % \item \textbf{(H1)} Practitioners exhibit a preference for \textit{visually interesting} projections when selecting DR projections for analysis.
    \item \textbf{(H1)} \textit{Visual interestingness} more strongly influences \textit{analytical preference} of DR projections than \textit{faithfulness} does. 
    \item \textbf{(H2)} The tendency observed in H1 becomes more pronounced when the color encoding of classes is provided.
    \item \textbf{(H3)} The tendency observed in H1 becomes more pronounced with shorter exposure time to stimuli.
    % when practitioners are exposed to projections in shorter time durations.
\end{itemize}
We thus demonstrate the substantial impact of System 1 on selecting DR projections for analysis (H1). 
We further show that System 1's influence intensifies with supporting factors such as color encoding and limited stimulus exposure time, aligned with prior observations in the literature (H2 and H3) \cite{morariu23tvcg, yunhai18tvcg}. Confirming these hypotheses establishes both the existence of bias in DR projection selection and its underlying mechanisms, thereby informing strategies for bias mitigation.

To achieve these objectives, we examine how the following two independent variables affect the visual interestingness and analytical preference of DR projections:
\begin{itemize}[leftmargin=15pt, itemsep=0pt, topsep=0.05cm]
    \item Color encoding of class labels (\textsc{ColorEncoding}): \textit{Monochrome} and \textit{Polychrome}
    \item Time duration for exposing stimuli (\textsc{ExposureTime}) : \textit{seven seconds} and \textit{15 seconds}
\end{itemize}
We select \textsc{ColorEncoding} because color-encoded class labels are commonly used for visualizing DR projections and significantly influence their perception \cite{morariu23tvcg, bibal16}. 
We also select \textsc{ExposureTime} as time constraints are known to amplify System 1 processing \cite{kahneman11thinking}.
We set exposure times of 7 and 15 seconds based on the time needed to read and interpret the metric scores in our pilot study.

Note that visual patterns of DR projections (e.g., class separability or the number of clusters) also affect their perception. 
For the completeness of our analysis, we also investigate their effect on visual interestingness in \autoref{sec:quant} (Analysis 2).

\subsection{Study Design}

Our experiment consists of two phases (\autoref{fig:userstudy}). In the first phase, we present participants with pairs of projections and ask them to select the one that are more visually interesting (H1). In the second phase, we present different participants with pairs of projections and ask them to select the one they prefer more for their analysis, while presenting faithfulness scores that favor projections with lower visual interest. 
We thus examine how practitioners' analytical preference for DR projections is affected by visual interestingness and faithfulness (H2). We control our independent variables (\textsc{ColorEncoding} and \textsc{ExposureTime}) for both phases.

\subsubsection{Phase 1: Visual Interestingness}

\label{sec:phase1}

We show participants pairs of projections and ask them to select a more visually interesting projection. 
% Participants perform pairwise comparisons by selecting the more visually interesting projection from presented pairs. 
Upon selection, the result feeds into an active ranking algorithm \cite{jamieson11neurips} that identifies optimal comparison pairs to efficiently converge on stable rankings with minimal iterations.
For each session, we ask participants to perform 50 trials of comparisons. Then, the active ranking algorithm outputs the final ranking based on the results of all trials. Participants complete four sessions, during which we distribute the combinations of independent variables to the sessions using a Latin square design.

\vspace{2.5pt}
\noindent\textbf{Procedure. }
One instructor manages all experiments. 
After participants make their consent and report demographics, the instructor details the objectives and tasks. 
We allow participants to freely ask questions during the introduction.
Then, participants go through four sessions. After the sessions, we conduct interviews to gather qualitative insights into their selections.

\vspace{2.5pt}
\noindent\textbf{Task. }
We provide participants with pairs of DR projections and ask them to determine which one is more visually interesting. We ask: \textit{``Given two projections, which projection catches your eye first?''}, following the definition of visual interestingness in Healey and Enns \cite{healey12tvcg}.
We employ pairwise comparison to obtain rankings that are unbiased by anchoring effects \cite{valdez18tvcg} and subjectivity inherent in Likert-scale-based evaluations \cite{south22cgf}.

\vspace{2.5pt}
\noindent\textbf{Distributing the combinations of independent variables.}
We have $2 \times 2 = 4$ combinations of \textsc{ColorEncoding} and \textsc{ExposureTime}. We distribute these conditions across participants using a Latin square design for counterbalancing, requiring participant counts in multiples of four. We recruit 16 participants in total.

\vspace{2.5pt}
\noindent\textbf{Stimuli. }
We generate 20 projections for each combination of \textsc{ColorEncoding} and \textsc{ExposureTime}, resulting in 80 projections. 
Here, we aim to maximize the diversity of visual patterns. 
To do so, we first select datasets with maximum variance in patterns using stratified sampling ~\cite{abbas19cgf, pandey16chi} from the existing set of 96 real-world datasets~\cite{jeon25tpami}.
Then, we generate various projections of the sampled datasets and again use stratified sampling to identify projections with diverse patterns (detailed procedure in Appendix A).

% . Please refer to Appendix A for the detailed procedure. 

\vspace{2.5pt}
\noindent\textbf{Determining the ranking of projections. }
% 우리는 매 세션이 끝나고, 앞선 50번의 선택을 바탕으로 20개의 projection에 대하여 visual interestingness rank를 매김. 이때 active ranking using pairwise comparison 알고리즘을 활용함.
After each session, we determine the visual interestingness rankings of the 20 projections based on the participants' 50 pairwise selections using an active ranking algorithm~\cite{jamieson11neurips}. Note that this algorithm also determines which pairs of projections to present to participants in each trial.

% \noindent\textbf{Participants. }
% We recruit 16 participants from four local universities (XX males and YY females, aged XX–YY [XX $\pm$ YY]). 
% We recruit participants who have experience in data analysis using scatterplots to align our experiments with real-world data analysis.
% XX of the participants report that they are experts in scatterplot-based visual analysis, while YY and ZZ report being intermediates and novices, respectively. 
% % ?? of the participants are undergraduates, ?? are graduate students, and ?? had just completed their Bachelor’s degree. 
% Also, we limit participants to those with low literacy of DR, including those who have prior experience with DR but have not engaged with it deeply.
% % DR에 대한 litercy가 높지 않은: DR을 사용은 해보았더라도 큰 고민 없이 사용해왔던,,
% Participants are compensated with the equivalent of \$7.

\vspace{2.5pt}
\noindent\textbf{Participants. }
We recruit 16 participants from four local universities (eight males and eight females, aged 21–33 [25.6 $\pm$ 3.5]). 
We recruit participants who have experience in data analysis using scatterplots to align our experiments with real-world data analysis.
% No participants report that they are experts in scatterplot-based visual analysis, while 11 and 5 report being intermediates and novices, respectively. 
No participants report being experts, while 11 and five report being intermediates and novices, respectively.
% ?? of the participants are undergraduates, ?? are graduate students, and ?? had just completed their Bachelor's degree. 
We limit participants to those with low literacy of DR, including those who have prior experience with DR but have not engaged with it deeply.
% DR에 대한 litercy가 높지 않은: DR을 사용은 해보았더라도 큰 고민 없이 사용해왔던,,
Participants are compensated with the equivalent of \$7.

\vspace{2.5pt}
\noindent\textbf{Apparatus. }
We conduct experiments over recorded Zoom calls. 
We develop a website where participants can view stimuli and make their selections with a mouse click or keyboard arrows. 
% We constrain participants to use a laptop or desktop screen to minimize the impact of the display on study results. 
We ask them to access the website and share their screens.

% \begin{figure}
%     \centering
%     \includegraphics[width=\linewidth]{figures/correlations.pdf}
%     \caption{The results of the analysis 1 (\autoref{sec:analysisone}). The analysis finds a significant effect of \textsc{ColorEncoding} on correlations between visual interestingness and analytical preference.}
%     \label{fig:analysisone}
% \end{figure}

% \begin{figure}
%     \centering
%     \includegraphics[width=0.965\linewidth]{figures/time_taken.pdf}
%     \caption{The time duration needed to select projections in our experiments. The figure shows that regardless of \textsc{ExposureTime} constraints, participants make a selection within approximately five seconds.}
%     \label{fig:duration}
% \end{figure}

\subsubsection{Phase 2: Analytical Preference}

Phase 2 shares most design choices and stimuli with Phase 1, except that we request analytical preferences: participants are guided to select projections that they wish to use for cluster analysis. 
We select cluster analysis for two reasons:
First, it is one of the most commonly applied analysis types for DR projections \cite{xia22tvcg, kwon18tvcg, jeon24tvcg}.
Second, cluster analysis is highly affected by our independent variable \textsc{ColorEncoding}.
Specifically, we ask participants: \textit{``Given two projections, which one do you prefer to use for analyzing the cluster structure of the original data?''}.
Additionally, we present faithfulness scores for the projections. These scores are artificially generated to favor projections with low visual interestingness, allowing us to examine which factor participants prioritize when visual interestingness and faithfulness conflict (H2).

% \noindent\textbf{Participants. }
% We recruit 16 participants from four local universities (XX males and YY females, aged XX–YY [XX $\pm$ YY]) following the same criteria as Phase 1. 
% XX of the participants report that they are experts in scatterplot-based visual analysis, while YY and ZZ report being intermediates and novices, respectively. 

\vspace{2.5pt}
\noindent\textbf{Participants. }
We recruit 16 participants from four local universities (11 males and five females, aged 23–30 [25.6 $\pm$ 2.1]) following the same criteria as Phase 1.
% 1 participant reports that the one is expert in scatterplot-based visual analysis, while 5 and 10 report being intermediates and novices, respectively.
One participant reports being an expert in scatterplot-based visual analysis, while five and 10 report being intermediate and novice practitioners, respectively.

% \noindent\textbf{Generating and presenting faithfulness scores.}
% For each pairwise comparison, we present five faithfulness scores for each projection. We use scores that are artificially generated, rather than the actual faithfulness scores of the projection, 
% to observe how practitioners select projections when visual interestingness and faithfulness scores contradicts; We intentionally generate scores that favors the projection with lower visual interesetingess over the higher one.

\vspace{2.5pt}
\noindent\textbf{Generating and presenting faithfulness scores.}
For each pairwise comparison, we present five pairs of faithfulness scores. 
We use artificially generated scores rather than actual faithfulness scores to examine how practitioners select projections when visual interestingness and faithfulness conflict.
Out of the five metrics, we assign higher scores to the less visually interesting projection on three or four randomly selected metrics. All scores randomly range within the [0, 1], consistent with the typical range of widely used DR metrics~\cite{hyeon23vis}.
% Also, we label each pair of faithfulness scores as metrics A through E. We use generic labels rather than actual metric names to avoid potential bias from participants' background knowledge.
% We use generic labels for each pair of faithfulness scores as metrics A through E, rather than actual metric names, to avoid potential bias from participants' background knowledge.
Also, to minimize potential bias from participants' background knowledge, we use generic labels (metrics A through E) for each pair of faithfulness scores, rather than actual metric names.

%% file: sections/05_results.tex
\section{Quantitative Analysis}

\label{sec:quant}

We present the results of the quantitative analysis of our user study.
% [[Results에 대한 세개를 종합하면 이렇다,, 가장 high-level (opening sentence)]]

\subsection{Analysis 1: Correlation between Rankings}

\label{sec:analysisone}

\noindent
\textbf{Objectives.} We examine the correlation between rankings of DR projections based on visual interestingness and analytical preference (H1). 
As we intentionally assign higher faithfulness scores to projections with lower visual interestingness, the correlation will ideally have the worst value if there is no bias from visual interestingness.
Conversely, positive correlations may occur when bias exists. 
We test this hypothesis and investigate how such correlations are affected by \textsc{ColorEncoding} and \textsc{ExposureTime} (H2, H3).

\vspace{2.5pt}
\noindent
\textbf{Analysis design.} We compute rankings of DR projections based on visual interestingness and analytical preference, then examine how 
such correlations vary across \textsc{ColorEncoding} and \textsc{ExposureTime}.
For each session, we obtain the rankings based on pairwise comparison results.
This yields 16 (participants) $\times$ 4 (sessions) = 64 rankings each for visual interestingness and analytical preference. 
With 2 (\textsc{ColorEncoding}) $\times$ 2 (\textsc{ExposureTime}) = 4 conditions and four dataset variations, we obtain 16 combinations in total.
Each combination contains four rankings for visual interestingness $\mathbf{V} = \{V_1, V_2, V_3, V_4\}$ and four rankings for analytical preference $\mathbf{A} = \{A_1, A_2, A_3, A_4\}$. 
For each combination, we compute the set of correlations between $\mathbf{V}$ and $\mathbf{A}$ as $\{\rho(V_i, A_i) \mid V_i \in \mathbf{V}, A_i \in \mathbf{A}\}$,
where $\rho$ denotes the Spearman correlation coefficient. 
We then analyze how these correlations vary across independent variables using two-way ANOVA with Tukey's HSD for post-hoc analysis.

\begin{figure}
    \centering
    \includegraphics[width=\linewidth]{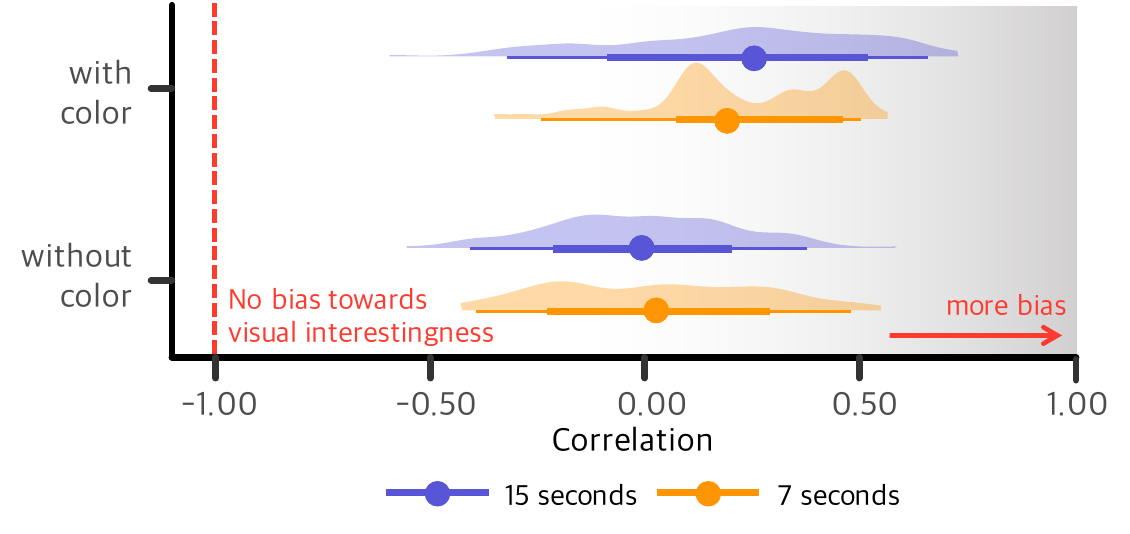}
    \vspace{-7mm}
    \caption{The correlations between visual interestingness and analytical preference in our experiment (\autoref{sec:analysisone}). We find that the correlations becomes significantly higher with color encoded class labels.}
    \label{fig:analysisone}
\end{figure}

\begin{figure}
    \centering
    \includegraphics[width=0.965\linewidth]{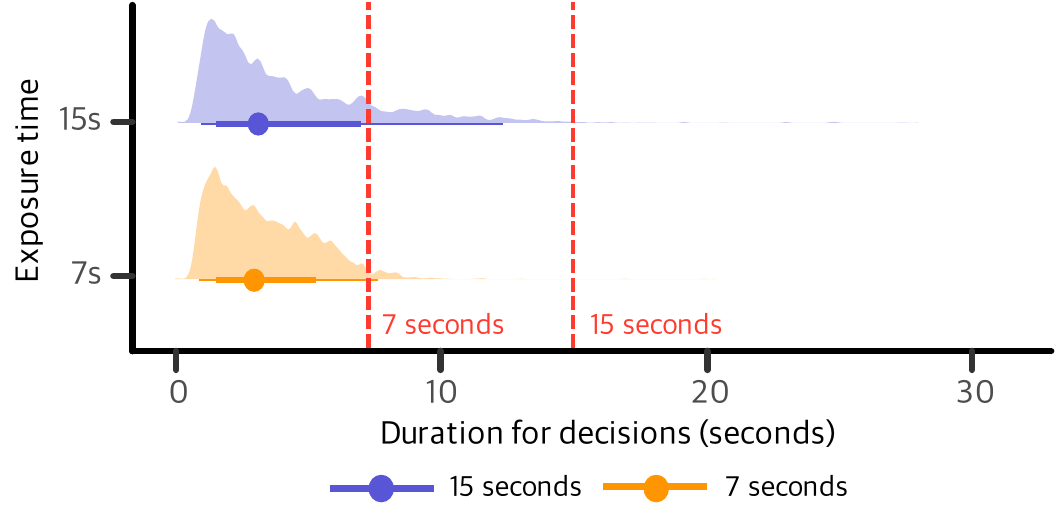}
    \vspace{-3.7mm}
    \caption{The time duration needed to select projections in our experiments (\autoref{sec:analysisone}). Regardless of \textsc{ExposureTime} constraints, participants make a selection within approximately five seconds.}
    \label{fig:duration}
\end{figure}

\vspace{2.5pt}
\noindent
\textbf{Results and discussions.} 
We find a significant effect on \textsc{ColorEncoding} ($F_{1, 336} = 65.10$, $p < .001$) but not on \textsc{ExposureTime} ($F_{1, 336} = 0.22$, $p = 0.64$), with no significant interaction effect ($F_{1, 336} = 0.46$, $p = 0.50$), confirming H1. 
For \textsc{ColorEncoding}, we find that correlations range around 0.25 (\autoref{fig:analysisone}). 
This confirms the ineffectiveness of adversarial depiction of faithfulness, supporting H2.
We observe that there is no significant effect for \textsc{ExposureTime} because participants complete their tasks within shorter durations---approximately five seconds on average---regardless of the time constraint (\autoref{fig:duration}).

\subsection{Analysis 2: Effect of Visual Features}

\label{sec:features}

% \begin{figure}
%     \centering
%     \includegraphics[width=0.965\linewidth]{figures/time_taken.pdf}
%     \caption{The time duration needed to select projections in our experiments. The figure shows that regardless of \textsc{ExposureTime} constraints, participants make a selection within approximately five seconds.}
%     \label{fig:duration}
% \end{figure}

\definecolor{applered}{RGB}{255,59,48}

\newcommand{\getopacity}[1]{%
    \pgfmathparse{#1 * 0.0333}%
    \pgfmathresult%
}

% 색상과 opacity를 적용하는 명령어
\newcommand{\cellshade}[1]{%
    \cellcolor{applered!#1!white}%
}

\begin{table*}[t]
    \centering
    \caption{The performance of linear regression models predicting visual interestingness from the features representing visual patterns of \textbf{polychrome} DR projections (number of clusters, cluster quality, class separation, and Scagnostics). 
    We investigate how the performance degrades as we remove each feature representing visual patterns. 
    The first row depicts the correlations of regression models, and the second row shows the performance decrement compared to the full model. 
    The opacity of cells represents the amount of decrement, with opacity mapped from 0.8 to 0 corresponding to 0\% to 15\%.
    % Unlike the results on monochrome projections, 
    \textit{Class separability} dominates the contribution of visual features in predicting visual interestingness, while other features show relatively small impact compared to their influence on monochrome projections.}
    \scalebox{0.87}{
    \begin{tabular}{rc|cccccccccccc}
    \toprule
             & \textit{full} & cluster \# & cluster qual. & class sep. & outlying & skewed & clumpy & sparse & striated & convex & skinny & stringy & monotonic \\ 
    \midrule 
       $R^2$  &  0.7784 & 
       \cellshade{4}0.7666 & 
       \cellshade{3}0.7689 & 
       \cellshade{64}0.5855 & 
       \cellshade{4}0.7663 & 
       \cellshade{9}0.7509 & 
       \cellshade{2}0.7722 & 
       \cellshade{14}0.7372 & 
       \cellshade{12}0.7432 & 
       \cellshade{5}0.7624 & 
       \cellshade{21}0.7163 & 
       \cellshade{4}0.7653 & 
       \cellshade{6}0.7605 \\
       Change  & $\cdot$ & 
       \cellshade{4}1.18\% & 
       \cellshade{3}0.95\% & 
       \cellshade{64}19.29\% & 
       \cellshade{4}1.21\% & 
       \cellshade{9}2.75\% & 
       \cellshade{2}0.61\% & 
       \cellshade{14}4.11\% & 
       \cellshade{12}3.52\% & 
       \cellshade{5}1.60\% & 
       \cellshade{21}6.20\% & 
       \cellshade{4}1.30\% & 
       \cellshade{6}1.79\% \\
    \bottomrule
    \end{tabular}
    }
    \label{tab:poly}
\end{table*}

\begin{table*}[t]
    \centering
    \caption{The performance of regression models predicting visual interestingness based on visual patterns of \textbf{monochrome} DR projections. 
    % We investigate how the performance degrades as we remove each feature representing visual patterns. 
    The table shares visual encoding and arrangement with \autoref{tab:poly}.
    \textit{Clumpy} shows the largest impact on visual interestingness, while other features exhibit smaller, comparable effects.}
    \scalebox{0.87}{
    \begin{tabular}{rc|cccccccccccc}
    \toprule
             & \textit{full} & cluster \# & cluster qual. & class sep. & outlying & skewed & clumpy & sparse & striated & convex & skinny & stringy & monotonic \\ 
    \midrule 
       $R^2$  &  0.7293 & 
       \cellshade{27}0.7062 & 
       \cellshade{24}0.7139 & 
       \cellshade{17}0.7294 & 
       \cellshade{20}0.7229 & 
       \cellshade{8}0.7499 & 
       \cellshade{44}0.6666 & 
       \cellshade{18}0.7265 & 
       \cellshade{14}0.7371 & 
       \cellshade{13}0.7381 & 
       \cellshade{14}0.7370 & 
       \cellshade{7}0.7532 & 
       \cellshade{19}0.7253 \\
       Change  & $\cdot$ & 
       \cellshade{27}8.09\% & 
       \cellshade{24}7.09\% & 
       \cellshade{17}5.07\% & 
       \cellshade{20}5.91\% & 
       \cellshade{8}2.40\% & 
       \cellshade{44}13.25\% & 
       \cellshade{18}5.45\% & 
       \cellshade{14}4.07\% & 
       \cellshade{13}3.94\% & 
       \cellshade{14}4.08\% & 
       \cellshade{7}1.98\% & 
       \cellshade{19}5.61\% \\
    \bottomrule
    \end{tabular}
    }
    \label{tab:mono}
\end{table*}

% : 알아보기 -> 어떤 feature가 biased selection을 유발하는가 

\noindent
\textbf{Objectives.} We want to understand how visual patterns beyond color encoding (e.g., class separability) of DR projections influence the visual interestingness of DR projections. 
% By doing so, we aim to understand which DR projections are susceptible to the bias towards visual interestingness and which patterns influence this bias, thereby informing our strategies to mitigate it. 
By doing so, we aim to identify not only which DR projections are susceptible to visual bias, but also which specific visual patterns drive this bias, thereby informing strategies to mitigate it.

% This investigation enables quantitative analysis of how critical each feature is in determining visual interestingness and analytical preference, providing empirical evidence for the relative importance of different scatterplot characteristics in human perception.

\vspace{2.5pt}
\noindent
\textbf{Analysis design.} 
We extract features representing visual patterns in DR projections and examine the extent to which these features affect visual interestingness. 
The \textit{number of clusters} and \textit{cluster quality} serve as our primary features, as prior work has demonstrated their significant influence on scatterplot perception \cite{jeon24tvcg, pandey16chi}.
For the number of clusters, we employ Gaussian Mixture Models across varying cluster numbers and select the configuration with the lowest Bayesian Information Criterion score, following 
% the methodology established in 
\cite{jeon24tvcg}.
We measure cluster quality using Silhouette scores due to their widespread use in cluster quality assessment in scatterplot 
% because it is widely used to examine cluster quality in scatterplots in literature 
\cite{bae25chi, joia11tvcg, kwon18tvcg}.
For polychrome projections, we additionally leverage \textit{class separability}, as this feature significantly influences scatterplot perception \cite{yunhai18tvcg, morariu23tvcg, bibal16}. 
We also incorporate Scagnostics \cite{wilkinson2005graph}, which substantially impacts how people perceive DR projections \cite{morariu23tvcg}.
% as it has been reported to substantially impact how people perceive DR projections 

We examine their influence on visual interestingness using an ablation study. We first train a regression model to predict the visual interestingness of DR projections using all extracted features. Subsequently, we train models with individual features removed and measure the resulting accuracy degradation relative to the full model, using this as a proxy for feature importance. 
% We repeat the same process by switching off the pair of features to examine their interplay. 
We use linear regression models as we have limited number of data points (which is 20), thus using other advanced model may suffer from data sparsity. We use $R^2$ as the target metric due to its interpretability \cite{cameron97econ}.

We apply this procedure independently to each set of 20 projections. For each set, we aggregate all participant trials and input them to the active ranking algorithm \cite{jamieson11neurips} to derive a consensus ranking of visual interestingness.
We assign visual interestingness scores (our target variable) using a linear transformation: projection with rank $r$ receives score $21 - r$, to align with the use of linear regression.

% such that the top-ranked projection obtains a score of 20.
% Note that the choice of scoring scale does not affect the robustness of our evaluation, as the decision tree structure underlying XGBoost is invariant to monotonic transformations of the target variable \cite{chen16kdd, leo22nips}.

\vspace{2.5pt}
\noindent
\textbf{Results and discussions.} 
\autoref{tab:poly} and \ref{tab:mono} depicts the results.
For polychrome projections, 
% (\autoref{tab:poly}), 
\textit{class separability} dominates the influence of visual features in predicting visual interestingness, with others showing substantially lower impact.
% Other features show substantially lower influence. 
In contrast, for monochrome projections, influences of features are more evenly distributed, where the impact of \textit{class separability} is substantially reduced compared to the case of polychrome projections. 
We also find that \textit{clumpy} exhibits the highest influence, with the \textit{number of clusters} and \textit{cluster quality} serve as runner-ups. 
These results indicate that when selecting DR projections, 
% \textit{class separability} and \textit{clumpy} in polychrome and monochrome projections, respectively, 
\textit{class separability} in polychrome projections and \textit{clumpy} in monochrome projections
intensify visual interestingness of practitioners, potentially stimulating System 1 processing.
% Such results indicate that when selecting DR projections, \textit{class separability} activates System 1 and triggers bias in polychrome projections, while \textit{clumpy} serves this role in monochrome projections.
We discuss the implications of such results in \autoref{sec:strategy}.

% , 그리고 monochrome projection의 clumpy가 visual interestingness를 더욱 강조한다. 이는 practitioners' system 1을 자극하여, faithfulness보다 visual interestingness를 priortize하도록 하는 bias를 일으킨다.

\section{Qualitative Results}

\label{sec:qualitative}

The following is our findings from the post-hoc interview study.

\vspace{2.5pt}
\noindent
\textbf{\textit{Finding 1.} Practitioners are biased towards visual interestingness over faithfulness in selecting DR projections. } 
We find qualitative evidence of the bias towards visual interestingness in selecting DR projections to analyze.
This tendency appeared regardless of the participants’ level of DR literacy. 
All participants report that they relied predominantly on the scatterplot’s visual characteristics when making their selections.
For example, P2 notes: "\textit{I relied much more on the visual part rather than the faithfulness metrics. When there was a conflict, I made decisions based more on the visual elements}."
Such findings provide additional evidence to confirm H1.

% Despite clearly explaining beforehand that the metrics represent the degree of faithfulness to the high-dimensional data—and that higher values indicate greater trustworthiness—most participants relied predominantly on the scatterplot’s visual characteristics when making their choices, largely ignoring the metrics.

\vspace{2.5pt}
\noindent
\textbf{\textit{Finding 2}. Practitioners exhibit stronger bias with color encoding and shorter exposure time. }
% → H3가 qualitative하게는 weak accept이다. 
We also find evidence that the bias towards visual interestingness intensifies with color-encoded labels and shorter exposure time. 16 participants report preferring polychrome projections over monochrome, noting that polychrome projections draw more visual attention. Moreover, we find that 
% when selecting projections that are analytically preferable, 
shorter exposure time compels practitioners to make selections based on the visual interestingness of projections. Ten participants report that they preferred more visually interesting projections when seven seconds are given, as they had limited time to read metric scores. These results provide weak support for H3.

\vspace{2.5pt}
\noindent
\textbf{\textit{Finding 3}. Practitioners are strongly influenced by clear separation of classes and clusters. }
Participants report that clearly separated classes and clusters draw their visual attention for polychrome and monochrome scatterplots, respectively, while selecting visually interesting projections.
The results agree with our quantitative findings (\autoref{sec:features}) that \textit{class separability} is an important feature determining the visual interestingness of polychrome projections, and that \textit{clumpy} serves the same role for monochrome projections.

% , that quantitative finding shows \textit{clumpy} as the strongest impact. 
% While the quantitative findings also identify \textit{class separability} and \textit{cluster quality} as important features affecting visual interestingness, they also reveal that other visual factors (e.g., \textit{clumpy} in monochrome projections) substantially influence perception of DR projections. 
% This discrepancy suggests that participants may not fully understand why they find certain projections visually interesting,
% % (e.g., \textit{clumpy} in monochrome projections not found in qualitative insights).
% as implicit factors like \textit{clumpy} in monochrome projections are not identified in their self-reported preferences.

% (\autoref{sec:features}), other visual factors (e.g., \textit{cluster numbers} in monochrome projections, or \textit{skinny} of polychrome projections) also affect participants' perception of scatterplots. Still, they failed to understand which factor they are visually attracted to. 

\vspace{2.5pt}
\noindent
\textbf{\textit{Finding 4}. Practitioners are unaware of their bias. }
We observe that all participants are unaware of their bias, regardless of their DR literacy level. Three participants also deny being biased. P2 states that they cannot agree that favoring visual interestingness over faithfulness scores should be considered bias.

%% file: sections/06_discussions.tex
\section{Discussions: Strategies to Mitigate Bias}

\label{sec:strategy}

Our findings inform the following strategies to mitigate bias towards visual interestingness in selecting DR projections to analyze. 

\vspace{2.5pt}
\noindent
\textbf{\textit{Strategy 1}: Deactivating System 1 in perceiving DR projections }
We recommend deactivating System 1 processing in perceiving DR projections by controlling visual factors in projections. 
As our findings reveal that (1) color encoding activates System 1 (\autoref{sec:analysisone}), and (2) class separability plays an important role in such activation (\autoref{sec:features} and \ref{sec:qualitative}), we suggest avoiding color encoding of class labels when depicting DR projections.
% to explain their relative faithfulness. 
One approach is to simply use monochrome representation. If displaying class labels is necessary (e.g., when showing DR performance for interactive labeling \cite{bernard18tvcg}), we recommend using shape encoding \cite{tseng25tvcg}.

\vspace{2.5pt}
\noindent
\textbf{\textit{Strategy 2}: Activating System 1 and 2 in reading faithfulness scores.}
We also suggest activating System 1 processing when reading faithfulness scores.
Our findings reveal that faithfulness scores are overlooked due to the visual interestingness of projections (\autoref{sec:analysisone}). These results indicate that faithfulness scores are less visually salient compared to projections. To reduce this gap, we recommend visually highlighting texts representing faithfulness scores \cite{strobelt16tvcg}, e.g., by assigning background color with high opacity to higher scores. Another plausible approach is to visually encode faithfulness scores, e.g., by depicting their distributions with uncertainty \cite{correll14tvcg} or leveraging word-scale visualizations \cite{goffin17tvcg}. These visual representations will not only increase System 1 engagement but also benefit System 2 processing, as interpreting visualizations requires additional cognitive load \cite{green08vast}.

\vspace{2.5pt}
\noindent
\textbf{\textit{Strategy 3}: Activating System 2 by enhancing DR literacy.}
Our findings indicate that projections with high clumpiness and well-separated clusters are perceived as visually interesting (\autoref{sec:features}). 
Regarding the fact that participants are unaware of their bias (\autoref{sec:qualitative}), such tendency may indicate that practitioners may erroneously favor DR techniques that exaggerate cluster structure, such as t-SNE \cite{maaten2008visualizing} and UMAP \cite{mcinnes2018umap}, as documented in the literature \cite{wattenberg2016how, hyeon25stopmisusing, kobak21naturebio, jeon25chi}. 
% This aligns with the prevalent misuse of t-SNE and UMAP identified by Jeon et al. \cite{hyeon25stopmisusing}.
To address this issue, we argue to invest community efforts in enhancing DR literacy.
This will lead practitioners to be more cautious in selecting DR projections to analyze (i.e., more influenced by System 2 processing). This direction will require practical efforts beyond academic papers, such as creating tutorials and approachable web articles \cite{wattenberg2016how, hyeon25stopmisusing}.

%% file: sections/07_conclusion.tex
\section{Conclusion}

In this study, we empirically demonstrate that practitioners are biased towards visual interestingness over faithfulness when selecting DR projections. Based on our findings, we recommend three strategies to mitigate this bias that deactivate System 1 processing and activate System 2 processing. In future work, we plan to investigate this bias more deeply and evaluate the effectiveness of our proposed strategies. Investigating how DR literacy impacts practitioners' perception of visual interestingness and faithfulness in DR projections would also be an interesting avenue to explore.

%% file: main.bbl
\begin{thebibliography}{10}
\renewcommand*{\sfdefault}{PTSansNarrow-TLF}

\bibitem{abbas19cgf}
M.~M. Abbas, M.~Aupetit, M.~Sedlmair, and H.~Bensmail.
\newblock Clustme: A visual quality measure for ranking monochrome scatterplots based on cluster patterns.
\newblock {\em Computer Graphics Forum}, 38(3):225--236, 2019. doi: \textsf{%
10\hspace{.1pt}\discretionary{.}{%
}{.}\hspace{.4pt}1111\discretionary{/}{%
}{/}cgf\hspace{.1pt}\discretionary{.}{%
}{.}\hspace{.4pt}13684}


\bibitem{bae25chi}
S.~S. Bae, T.~Fujiwara, C.~Tseng, and D.~Szafir.
\newblock Uncovering how scatterplot features skew visual class separation.
\newblock In {\em ACM CHI}, 2025.

\bibitem{bernard18tvcg}
J.~Bernard, M.~Hutter, M.~Zeppelzauer, D.~Fellner, and M.~Sedlmair.
\newblock Comparing visual-interactive labeling with active learning: An experimental study.
\newblock {\em IEEE Transactions on Visualization and Computer Graphics}, 24(1):298--308, 2018. doi: \textsf{%
10\hspace{.1pt}\discretionary{.}{%
}{.}\hspace{.4pt}1109\discretionary{/}{%
}{/}TVCG\hspace{.1pt}\discretionary{.}{%
}{.}\hspace{.4pt}2017\hspace{.1pt}\discretionary{.}{%
}{.}\hspace{.4pt}2744818}


\bibitem{bibal16}
A.~Bibal and B.~Frénay.
\newblock Learning interpretability for visualizations using adapted cox models through a user experiment, 2016.

\bibitem{cashman25arxiv}
D.~Cashman, M.~Keller, H.~Jeon, B.~C. Kwon, and Q.~Wang.
\newblock A critical analysis of the usage of dimensionality reduction in four domains.
\newblock {\em IEEE Transactions on Visualization and Computer Graphics}, pp. 1--20, 2025. doi: \textsf{%
10\hspace{.1pt}\discretionary{.}{%
}{.}\hspace{.4pt}1109\discretionary{/}{%
}{/}TVCG\hspace{.1pt}\discretionary{.}{%
}{.}\hspace{.4pt}2025\hspace{.1pt}\discretionary{.}{%
}{.}\hspace{.4pt}3567989}


\bibitem{cameron97econ}
A.~{Colin Cameron} and F.~A. Windmeijer.
\newblock An r-squared measure of goodness of fit for some common nonlinear regression models.
\newblock {\em Journal of Econometrics}, 77(2):329--342, 1997. doi: \textsf{%
10\hspace{.1pt}\discretionary{.}{%
}{.}\hspace{.4pt}1016\discretionary{/}{%
}{/}S0304\discretionary{%
}{-}{-}4076\discretionary{%
}{(}{(}96\discretionary{)}{%
}{)}01818\discretionary{%
}{-}{-}0}


\bibitem{correll14tvcg}
M.~Correll and M.~Gleicher.
\newblock Error bars considered harmful: Exploring alternate encodings for mean and error.
\newblock {\em IEEE Transactions on Visualization and Computer Graphics}, 20(12):2142--2151, 2014. doi: \textsf{%
10\hspace{.1pt}\discretionary{.}{%
}{.}\hspace{.4pt}1109\discretionary{/}{%
}{/}TVCG\hspace{.1pt}\discretionary{.}{%
}{.}\hspace{.4pt}2014\hspace{.1pt}\discretionary{.}{%
}{.}\hspace{.4pt}2346298}


\bibitem{friedman87}
J.~H. Friedman.
\newblock Exploratory projection pursuit.
\newblock {\em Journal of the American statistical association}, 82(397):249--266, 1987.

\bibitem{goffin17tvcg}
P.~Goffin, J.~Boy, W.~Willett, and P.~Isenberg.
\newblock An exploratory study of word-scale graphics in data-rich text documents.
\newblock {\em IEEE Transactions on Visualization and Computer Graphics}, 23(10):2275--2287, 2017. doi: \textsf{%
10\hspace{.1pt}\discretionary{.}{%
}{.}\hspace{.4pt}1109\discretionary{/}{%
}{/}TVCG\hspace{.1pt}\discretionary{.}{%
}{.}\hspace{.4pt}2016\hspace{.1pt}\discretionary{.}{%
}{.}\hspace{.4pt}2618797}


\bibitem{green08vast}
T.~M. Green, W.~Ribarsky, and B.~Fisher.
\newblock Visual analytics for complex concepts using a human cognition model.
\newblock In {\em 2008 IEEE Symposium on Visual Analytics Science and Technology}, pp. 91--98, 2008. doi: \textsf{%
10\hspace{.1pt}\discretionary{.}{%
}{.}\hspace{.4pt}1109\discretionary{/}{%
}{/}VAST\hspace{.1pt}\discretionary{.}{%
}{.}\hspace{.4pt}2008\hspace{.1pt}\discretionary{.}{%
}{.}\hspace{.4pt}4677361}


\bibitem{healey12tvcg}
C.~Healey and J.~Enns.
\newblock Attention and visual memory in visualization and computer graphics.
\newblock {\em IEEE Transactions on Visualization and Computer Graphics}, 18(7):1170--1188, 2012. doi: \textsf{%
10\hspace{.1pt}\discretionary{.}{%
}{.}\hspace{.4pt}1109\discretionary{/}{%
}{/}TVCG\hspace{.1pt}\discretionary{.}{%
}{.}\hspace{.4pt}2011\hspace{.1pt}\discretionary{.}{%
}{.}\hspace{.4pt}127}


\bibitem{jamieson11neurips}
K.~G. Jamieson and R.~Nowak.
\newblock Active ranking using pairwise comparisons.
\newblock {\em Advances in neural information processing systems}, 24, 2011.

\bibitem{jeon25tpami}
H.~Jeon, M.~Aupetit, D.~Shin, A.~Cho, S.~Park, and J.~Seo.
\newblock Measuring the validity of clustering validation datasets.
\newblock {\em IEEE Transactions on Pattern Analysis and Machine Intelligence}, 47(6):5045--5058, 2025. doi: \textsf{%
10\hspace{.1pt}\discretionary{.}{%
}{.}\hspace{.4pt}1109\discretionary{/}{%
}{/}TPAMI\hspace{.1pt}\discretionary{.}{%
}{.}\hspace{.4pt}2025\hspace{.1pt}\discretionary{.}{%
}{.}\hspace{.4pt}3548011}


\bibitem{hyeon23vis}
H.~Jeon, A.~Cho, J.~Jang, S.~Lee, J.~Hyun, H.-K. Ko, J.~Jo, and J.~Seo.
\newblock Zadu: A python library for evaluating the reliability of dimensionality reduction embeddings.
\newblock In {\em 2023 IEEE Visualization and Visual Analytics (VIS)}, pp. 196--200. doi: \textsf{%
10\hspace{.1pt}\discretionary{.}{%
}{.}\hspace{.4pt}1109\discretionary{/}{%
}{/}VIS54172\hspace{.1pt}\discretionary{.}{%
}{.}\hspace{.4pt}2023\hspace{.1pt}\discretionary{.}{%
}{.}\hspace{.4pt}00048}


\bibitem{jeon25chi}
H.~Jeon, H.~Lee, Y.-H. Kuo, T.~Yang, D.~Archambault, S.~Ko, T.~Fujiwara, K.-L. Ma, and J.~Seo.
\newblock Unveiling high-dimensional backstage: A survey for reliable visual analytics with dimensionality reduction.
\newblock In {\em Proceedings of the 2025 CHI Conference on Human Factors in Computing Systems}, CHI '25, 2025. doi: \textsf{%
10\hspace{.1pt}\discretionary{.}{%
}{.}\hspace{.4pt}1145\discretionary{/}{%
}{/}3706598\hspace{.1pt}\discretionary{.}{%
}{.}\hspace{.4pt}3713551}


\bibitem{hyeon25stopmisusing}
H.~Jeon, J.~Park, S.~Shin, and J.~Seo.
\newblock Stop misusing t-sne and umap for visual analytics, 2025.

\bibitem{jeon24tvcg}
H.~Jeon, G.~J. Quadri, H.~Lee, P.~Rosen, D.~A. Szafir, and J.~Seo.
\newblock Clams: A cluster ambiguity measure for estimating perceptual variability in visual clustering.
\newblock {\em IEEE Transactions on Visualization and Computer Graphics}, 30(1):770--780, 2024. doi: \textsf{%
10\hspace{.1pt}\discretionary{.}{%
}{.}\hspace{.4pt}1109\discretionary{/}{%
}{/}TVCG\hspace{.1pt}\discretionary{.}{%
}{.}\hspace{.4pt}2023\hspace{.1pt}\discretionary{.}{%
}{.}\hspace{.4pt}3327201}


\bibitem{joia11tvcg}
P.~Joia, D.~Coimbra, J.~A. Cuminato, F.~V. Paulovich, and L.~G. Nonato.
\newblock Local affine multidimensional projection.
\newblock {\em IEEE Transactions on Visualization and Computer Graphics}, 17(12):2563--2571, 2011. doi: \textsf{%
10\hspace{.1pt}\discretionary{.}{%
}{.}\hspace{.4pt}1109\discretionary{/}{%
}{/}TVCG\hspace{.1pt}\discretionary{.}{%
}{.}\hspace{.4pt}2011\hspace{.1pt}\discretionary{.}{%
}{.}\hspace{.4pt}220}


\bibitem{kahneman11thinking}
D.~Kahneman.
\newblock Thinking, fast and slow.
\newblock 2011.

\bibitem{kobak21naturebio}
D.~Kobak and G.~C. Linderman.
\newblock Initialization is critical for preserving global data structure in both t-sne and umap.
\newblock {\em Nature biotechnology}, 39(2):156--157, 2021. doi: \textsf{%
10\hspace{.1pt}\discretionary{.}{%
}{.}\hspace{.4pt}1038\discretionary{/}{%
}{/}s41587\discretionary{%
}{-}{-}020\discretionary{%
}{-}{-}00809\discretionary{%
}{-}{-}z}


\bibitem{kwon18tvcg}
B.~C. Kwon, B.~Eysenbach, J.~Verma, K.~Ng, C.~De~Filippi, W.~F. Stewart, and A.~Perer.
\newblock Clustervision: Visual supervision of unsupervised clustering.
\newblock {\em IEEE Transactions on Visualization and Computer Graphics}, 24(1):142--151, 2018. doi: \textsf{%
10\hspace{.1pt}\discretionary{.}{%
}{.}\hspace{.4pt}1109\discretionary{/}{%
}{/}TVCG\hspace{.1pt}\discretionary{.}{%
}{.}\hspace{.4pt}2017\hspace{.1pt}\discretionary{.}{%
}{.}\hspace{.4pt}2745085}


\bibitem{maaten2008visualizing}
L.~v.~d. Maaten and G.~Hinton.
\newblock Visualizing data using t-sne.
\newblock {\em Journal of machine learning research}, 9(Nov):2579--2605, 2008.

\bibitem{mcinnes2018umap}
L.~McInnes, J.~Healy, and J.~Melville.
\newblock Umap: Uniform manifold approximation and projection for dimension reduction.
\newblock {\em arXiv preprint arXiv:1802.03426}, 2018.

\bibitem{morariu23tvcg}
C.~Morariu, A.~Bibal, R.~Cutura, B.~Frénay, and M.~Sedlmair.
\newblock Predicting user preferences of dimensionality reduction embedding quality.
\newblock {\em IEEE Transactions on Visualization and Computer Graphics}, 29(1):745--755, 2023. doi: \textsf{%
10\hspace{.1pt}\discretionary{.}{%
}{.}\hspace{.4pt}1109\discretionary{/}{%
}{/}TVCG\hspace{.1pt}\discretionary{.}{%
}{.}\hspace{.4pt}2022\hspace{.1pt}\discretionary{.}{%
}{.}\hspace{.4pt}3209449}


\bibitem{nguyen13pvis}
Q.~Nguyen, P.~Eades, and S.-H. Hong.
\newblock On the faithfulness of graph visualizations.
\newblock In {\em 2013 IEEE Pacific Visualization Symposium (PacificVis)}, pp. 209--216, 2013. doi: \textsf{%
10\hspace{.1pt}\discretionary{.}{%
}{.}\hspace{.4pt}1109\discretionary{/}{%
}{/}PacificVis\hspace{.1pt}\discretionary{.}{%
}{.}\hspace{.4pt}2013\hspace{.1pt}\discretionary{.}{%
}{.}\hspace{.4pt}6596147}


\bibitem{nonato19tvcg}
L.~G. Nonato and M.~Aupetit.
\newblock Multidimensional projection for visual analytics: Linking techniques with distortions, tasks, and layout enrichment.
\newblock {\em IEEE Transactions on Visualization and Computer Graphics}, 25(8):2650--2673, 2019. doi: \textsf{%
10\hspace{.1pt}\discretionary{.}{%
}{.}\hspace{.4pt}1109\discretionary{/}{%
}{/}TVCG\hspace{.1pt}\discretionary{.}{%
}{.}\hspace{.4pt}2018\hspace{.1pt}\discretionary{.}{%
}{.}\hspace{.4pt}2846735}


\bibitem{pandey16chi}
A.~V. Pandey, J.~Krause, C.~Felix, J.~Boy, and E.~Bertini.
\newblock Towards understanding human similarity perception in the analysis of large sets of scatter plots.
\newblock In {\em Proceedings of the 2016 CHI Conference on Human Factors in Computing Systems}, CHI '16, p. 3659–3669, 2016. doi: \textsf{%
10\hspace{.1pt}\discretionary{.}{%
}{.}\hspace{.4pt}1145\discretionary{/}{%
}{/}2858036\hspace{.1pt}\discretionary{.}{%
}{.}\hspace{.4pt}2858155}


\bibitem{seo05infovis}
J.~Seo and B.~Shneiderman.
\newblock A rank-by-feature framework for interactive exploration of multidimensional data.
\newblock {\em Information visualization}, 4(2):96--113, 2005.

\bibitem{south22cgf}
L.~South, D.~Saffo, O.~Vitek, C.~Dunne, and M.~A. Borkin.
\newblock Effective use of likert scales in visualization evaluations: A systematic review.
\newblock {\em Computer Graphics Forum}, 41(3):43--55, 2022. doi: \textsf{%
10\hspace{.1pt}\discretionary{.}{%
}{.}\hspace{.4pt}1111\discretionary{/}{%
}{/}cgf\hspace{.1pt}\discretionary{.}{%
}{.}\hspace{.4pt}14521}


\bibitem{strobelt16tvcg}
H.~Strobelt, D.~Oelke, B.~C. Kwon, T.~Schreck, and H.~Pfister.
\newblock Guidelines for effective usage of text highlighting techniques.
\newblock {\em IEEE Transactions on Visualization and Computer Graphics}, 22(1):489--498, 2016. doi: \textsf{%
10\hspace{.1pt}\discretionary{.}{%
}{.}\hspace{.4pt}1109\discretionary{/}{%
}{/}TVCG\hspace{.1pt}\discretionary{.}{%
}{.}\hspace{.4pt}2015\hspace{.1pt}\discretionary{.}{%
}{.}\hspace{.4pt}2467759}


\bibitem{tseng25tvcg}
C.~Tseng, A.~Z. Wang, G.~J. Quadri, and D.~A. Szafir.
\newblock Shape it up: An empirically grounded approach for designing shape palettes.
\newblock {\em IEEE Transactions on Visualization and Computer Graphics}, 31(1):349--359, 2025. doi: \textsf{%
10\hspace{.1pt}\discretionary{.}{%
}{.}\hspace{.4pt}1109\discretionary{/}{%
}{/}TVCG\hspace{.1pt}\discretionary{.}{%
}{.}\hspace{.4pt}2024\hspace{.1pt}\discretionary{.}{%
}{.}\hspace{.4pt}3456385}


\bibitem{tversky74science}
A.~Tversky and D.~Kahneman.
\newblock Judgment under uncertainty: Heuristics and biases: Biases in judgments reveal some heuristics of thinking under uncertainty.
\newblock {\em science}, 185(4157):1124--1131, 1974.

\bibitem{valdez18tvcg}
A.~C. Valdez, M.~Ziefle, and M.~Sedlmair.
\newblock Priming and anchoring effects in visualization.
\newblock {\em IEEE Transactions on Visualization and Computer Graphics}, 24(1):584--594, 2018. doi: \textsf{%
10\hspace{.1pt}\discretionary{.}{%
}{.}\hspace{.4pt}1109\discretionary{/}{%
}{/}TVCG\hspace{.1pt}\discretionary{.}{%
}{.}\hspace{.4pt}2017\hspace{.1pt}\discretionary{.}{%
}{.}\hspace{.4pt}2744138}


\bibitem{venna06nn}
J.~Venna and S.~Kaski.
\newblock Local multidimensional scaling.
\newblock {\em Neural Networks}, 19(6):889--899, 2006.
\newblock Advances in Self Organising Maps - WSOM’05. doi: \textsf{%
10\hspace{.1pt}\discretionary{.}{%
}{.}\hspace{.4pt}1016\discretionary{/}{%
}{/}j\hspace{.1pt}\discretionary{.}{%
}{.}\hspace{.4pt}neunet\hspace{.1pt}\discretionary{.}{%
}{.}\hspace{.4pt}2006\hspace{.1pt}\discretionary{.}{%
}{.}\hspace{.4pt}05\hspace{.1pt}\discretionary{.}{%
}{.}\hspace{.4pt}014}


\bibitem{yunhai18tvcg}
Y.~Wang, K.~Feng, X.~Chu, J.~Zhang, C.-W. Fu, M.~Sedlmair, X.~Yu, and B.~Chen.
\newblock A perception-driven approach to supervised dimensionality reduction for visualization.
\newblock {\em IEEE Transactions on Visualization and Computer Graphics}, 24(5):1828--1840, 2018. doi: \textsf{%
10\hspace{.1pt}\discretionary{.}{%
}{.}\hspace{.4pt}1109\discretionary{/}{%
}{/}TVCG\hspace{.1pt}\discretionary{.}{%
}{.}\hspace{.4pt}2017\hspace{.1pt}\discretionary{.}{%
}{.}\hspace{.4pt}2701829}


\bibitem{wattenberg2016how}
M.~Wattenberg, F.~Viégas, and I.~Johnson.
\newblock How to use t-sne effectively.
\newblock {\em Distill}, 2016. doi: \textsf{%
10\hspace{.1pt}\discretionary{.}{%
}{.}\hspace{.4pt}23915\discretionary{/}{%
}{/}distill\hspace{.1pt}\discretionary{.}{%
}{.}\hspace{.4pt}00002}


\bibitem{wilkinson2005graph}
L.~Wilkinson, A.~Anand, and R.~Grossman.
\newblock Graph-theoretic scagnostics.
\newblock In {\em Information visualization, IEEE symposium on}, pp. 21--21. IEEE Computer Society, 2005.

\bibitem{xia22tvcg}
J.~Xia, Y.~Zhang, J.~Song, Y.~Chen, Y.~Wang, and S.~Liu.
\newblock Revisiting dimensionality reduction techniques for visual cluster analysis: An empirical study.
\newblock {\em IEEE Transactions on Visualization and Computer Graphics}, 28(1):529--539, 2022. doi: \textsf{%
10\hspace{.1pt}\discretionary{.}{%
}{.}\hspace{.4pt}1109\discretionary{/}{%
}{/}TVCG\hspace{.1pt}\discretionary{.}{%
}{.}\hspace{.4pt}2021\hspace{.1pt}\discretionary{.}{%
}{.}\hspace{.4pt}3114694}


\end{thebibliography}
